\documentclass[preprint,aps,showpacs,superscriptaddress,floatfix]{revtex4-1}
\usepackage{graphicx}
\usepackage{dcolumn}
\usepackage{bm}
\usepackage{color}

\begin{document}

\preprint{APS/123-QED}

\title{Three-dimensional structure of Mach cones in monolayer complex plasma crystals}

\author{L. Cou\"edel}
\affiliation{Aix-Marseille Univ./CNRS, Laboratoire PIIM, 13397 Marseille Cedex 20, France}
\email{lenaic.couedel@univ-amu.fr}


\author{D. Samsonov\footnote{On September 25, 2012, our dear friend and colleague, Dmitry Samsonov, passed away. He is remembered by us  
for his excellent research, his great sense of humour, his very individual style of driving and towards 
the latter years for the courage and braveness with which he fought his terrible illness. We have not only lost one of the pioneers 
of complex plasma physics but a remarkable person.
}}
\author{C. Durniak}
\affiliation{Department of Electrical Engineering and Electronics, The
University of Liverpool, Liverpool, L69 3GJ, United Kingdom}

\author{S. Zhdanov}
\author{H.M. Thomas}
\author{G.E. Morfill}
\affiliation{Max Planck Institute for Extraterrestrial Physics, 85741 Garching, Germany}

\author{C. Arnas}
\affiliation{Aix-Marseille Univ./CNRS, Laboratoire PIIM, 13397 Marseille Cedex 20, France}

\date{\today}

\begin{abstract}

Structure of Mach cones in a crystalline complex plasma has been
studied experimentally using an intensity sensitive imaging, which
resolved particle motion in three dimensions. This revealed a
previously unknown out-of-plane cone structure, which appeared due to
excitation of the vertical wave mode.  The complex plasma consisted of
micron sized particles forming a monolayer in a plasma sheath of a gas
discharge. Fast particles, spontaneously moving under the monolayer,
created Mach cones with multiple structures. The in-plane cone
structure was due to compressional and shear lattice waves.
\end{abstract}


\maketitle

Complex (or dusty) plasmas are weakly ionised gases containing dust.
Due to absorption of ambient electrons and ions of the plasma, dust
particles acquire significant negative electric charges
\cite{Bouchoule1999,Morfill2009,Vladimirov2005,Arnas2001,Shukla2002}. They
interact strongly with the plasma and with each other and they can
form strongly coupled systems analogous to colloids. Dust particles
are often confined in the sheath region of electrical discharges,
where the electrostatic force is strong enough to compensate their
weight. The sheath region confines the particles strongly in the
vertical direction and they can form monolayer 
crystals \cite{Hebner2002,Samsonov2005}. 
\textcolor{black}{These systems extend  in the 'third' vertical direction to their finite width 
which depends on the strength of vertical confinement \cite{Samsonov2005}. A vertical ``out-of-plane''  
particle motion is allowed making the dust system geometry  three-dimensional. However, the 
magnitude of the vertical displacement is very small compared to the interparticle distance 
(tens of micrometers compared to hundred of micrometers). For this reason, monolayer complex 
plasma crystals are often referred as two-dimensional or quasi-two-dimensional (quasi-2D) complex plasma crystals \cite{Couedel2009a}. 
A monolayer plasma crystal has not to be confused with a dust multilayer system \cite{Pieper1996}.}

If a body moves in a wave sustaining medium, it creates a
 disturbance, also known as a wake. V-shaped 
 wakes, also known as Mach cones, are created behind a body moving supersonically. 
These cones are observed behind boats, supersonic jets, in fluid filled boreholes,
and in complex plasmas.  The structure of Mach cones is determined by
the wave modes that exist in the medium, with each mode having a
distinct contribution.  \textcolor{black}{Quasi-2D} lattices sustain two \textit{in-plane} wave
modes with acoustic dispersion.  One of them is compressional
(longitudinal), the other is shear (transverse) \cite{Wang2001}. Compressional cones
have been observed in a \textcolor{black}{quasi-2D} complex plasma using spontaneously
accelerated particles \cite{Samsonov1999b,Samsonov2000,Dubin2000}.
They often have a multiple structure with a compressional cone
followed by a rarefactional and then possibly by another
compressional \cite{Schwabe2011}. The cone angle $\mu$ is determined by the Mach cone
rule $\sin \mu = c/V$, where $V$ the speed of the supersonic particle
and $c$ is the speed of the wave that comprises the cone. If the waves are
dispersive (e.g. water waves) or change their speed as they propagate
in an inhomogeneous medium, the Mach cone angle will also change
\cite{Zhdanov2004}.  Wakes produced by shear waves have been observed
experimentally using laser excitation \cite{Melzer2000,Nosenko2002}.
Since the shear mode is about 4-5 times slower than the compressional one,
the shear cone fits within the compressional cones, so that both
structures can be observed simultaneously. Shear cones have been
generated by supersonic dislocations in stressed complex plasma
crystals \cite{Nosenko2007}.

Since the strength of the vertical confinement in \textcolor{black}{quasi-2D} complex plasma
systems is finite, there is a third fundamental wave mode: vertical
transverse wave (VTW).  It is associated with propagating
\textit{out-of-plane} or vertical oscillations
\cite{Couedel2010,Couedel2009a,Qiao2003} and it has an optical dispersion.  It
is then natural to ask whether a vertical Mach cone can be produced by
VTW. Optical wave packets propagate with the group velocity, which is,
in case of VTW, usually slightly lower than the shear wave speed
\cite{Samsonov2005}. Thus it should be possible to observe
compressional, shear and vertical cone structures at the same time.
Here we report a study of the 3D structure of Mach cones
using an intensity sensitive imaging technique
\cite{Samsonov2005,Couedel2009a}, which measures vertical
displacements of particles in a monolayer lattice. This method has
allowed us to resolve compressional, shear, and vertical cone
structures.

The experiments were performed in two very similar capacitively
coupled radio frequency (RF) glow discharge chambers at 13.56 MHz, one
\cite{Couedel2009a} at Max Planck Institute for Extraterrestrial
Physics in Garching (experiment I) and the other \cite{Durniak2011} at
the department of Electrical Engineering and Electronics at the
University of Liverpool (experiment II). 
The experimental parameters are listed in Table~\ref{tab1}. The
chambers were initially pumped down using turbo-molecular pumps. The
working gas pressure was maintained by a small argon flow set to
minimise the disturbance to the lattice.  The RF power was applied
between the lower electrode ($\approx$200~mm in diameter) and the
grounded chamber. Due to electrode asymmetry, the powered electrode
acquired a negative self-bias voltage. Typical plasma parameters have
been measured with a Langmuir probe in the bulk discharge at a
pressure of 0.66~Pa and discharge power of 20~W \cite{Nosenko2009}
yielding the electron temperature $T_e=2.5$~eV and the electron
density $n_e=2\times10^9$~cm$^{-3}$.  A monolayer particle suspension
(50--60~mm in diameter) was formed by injecting melamine formaldehyde
microspheres (9.19$\pm$0.14~$\mu$m in diameter, with a mass of
6.1$\times$10$^{-13}$~kg) into the plasma sheath above the lower
electrode.

\begin{table*}[htbp] 
\caption{Parameters used in the experiments and simulation.}
\label{tab1}
\begin{tabular}{llcccc}
    \hline
    \hline
     Parameter                              & Symbol                   & Unit                  & Experiment I   & Experiment II & Simulation\\
     \hline              
     argon pressure                         & $p$                      & (Pa)                  & 0.66           & 1.45          & -- \\
     RF power                               & $P$                      & (W)                   & 15             & 3             & -- \\
     camera frame rate                      &                          & (fps)                 & 250            & 125           & -- \\ 
     average interparticle distance         & $a$                      & ($\mu$m)              & 556            & 576           & 425 \\ 
     longitudinal dust-lattice wave speed   & $C_L$                    & (mm/s)                & 31.7 	        & 23.4          & 38.1 \\ 
     screening parameter                    & $\kappa=a/\lambda_D$     &                       & 1.23           & 1.40          & 0.94 \\ 
     Debye length                           & $\lambda_D$              & ($\mu$m)              & 452            & 410           & 452 \\ 
     particle charge in electron charges    & $Z$                    & ($e$)                 & 18000         & 14800        & 18000 \\ 
     vertical resonance frequency           & $f_v$                    & (Hz)                  & 26.0 	        & 16.5          & 23.9\\ 
     speed of perturbing particle           & $V$                  & (mm/s)                & 40.2 	        & 19.2          & 40.0 \\
     longitudinal Mach number               & $M_L=V/C_L$          &                       & 1.2            & 0.8           & 1.05 \\
     longitudinal Mach angle                & $\mu_L=\arcsin(1/M_L)$   & (deg.)                & 56             & --            & 72.5\\
     measured horizontal damping            & $\gamma_h$               & (s$^{-1}$)            & 0.83 $\pm$0.21 & 1.57 $\pm$0.63& -- \\
     theoretical damping (Epstein)          & $\gamma_{th}$            & (s$^{-1}$)            & 0.79           & 1.73          & 1.0 \\
   \hline
   \hline
\end{tabular}
\end{table*}

The microparticles were illuminated by a thin horizontal laser sheet,
which had a Gaussian profile in the vertical direction and a uniform
profile in the horizontal direction.  The sheet thickness was nearly
constant across the crystal.  The particles were imaged through the
top window by a Photron FASTCAM high speed camera. The horizontal
coordinates $x$ and $y$ as well as the velocity components $v_x$ and
$v_y$ of individual particles were then measured with sub-pixel
resolution using standard particle tracking techniques
\cite{Rogers2007,Feng2011,Ivanov2007}. An additional side-view camera
was used to verify that our experiments were carried out with a single
main layer of particles. In order to determine the relative vertical
positions $z$ and velocities $v_z$ of individual particles, the laser
sheet was set slightly above the particle layer, so that the position
of the intensity maximum was $100~\mu$m to $200~\mu$m higher than the
average levitation height of the grains. The intensity of light
scattered by the particles depended on their vertical displacement
from the average height.  The vertical velocity was calculated from
the change of particle intensity in two consecutive frames
\cite{Samsonov2005,Couedel2009a}.

The complex plasma lattices used in our experiments have been
characterised by measuring their main parameters, which are shown in
Table~\ref{tab1}.  The charge and the screening parameter in both
experiments were determined from the long wavelength phonon spectra of
the thermally excited lattice oscillations.  The spectra have been
analysed yielding the longitudinal and transverse dust-lattice wave
speeds with the method of Ref.~\cite{Nunomura2002}.  The particle
charge and the screening parameter were then calculated using the
formulae of Ref.~\cite{Piel2002}. The interparticle distances
were measured directly from the lattice images, and the Debye
length was calculated. We have also measured the vertical resonance
frequency, which characterises the strength of the vertical
confinement.

Mach cones were created using the experimental technique of
Refs.~\cite{Samsonov1999b,Samsonov2000}. A few heavy particles
(which are often found in many experiments) moved under the main
lattice layer. They were spontaneously accelerated, most likely by the
wake fields of particles in the main layer \cite{Schweigert2002}, if
the RF power was high enough. Their speeds were nearly constant,
limited by the neutral gas friction (Epstein drag), and their
trajectories were most often straight lines. They produced Mach 
cones due to electrostatic interaction with the
crystal lattice above.

In order to understand the mechanisms of lattice excitation we
reproduced the Mach cones with the molecular dynamics simulation code
of Ref.~\cite{durniak10:_simulation}.  It solved the equations of
motion of 3000 microparticles (with the same mass as in the
experiments) interacting with each other via a Yukawa potential. They
were confined in a parabolic potential well, strongly in the vertical
and weakly in the horizontal directions, forming a monolayer.
Parameters of the simulation are listed in Table~\ref{tab1}.  The
lattice was excited by a 3D repulsive spot force field moving at
40~mm/s along the x-axis.  The force field was similar to that of a
charged particle.

\begin{figure*}[htbp]
\centering
\includegraphics[width=\linewidth,angle=0]{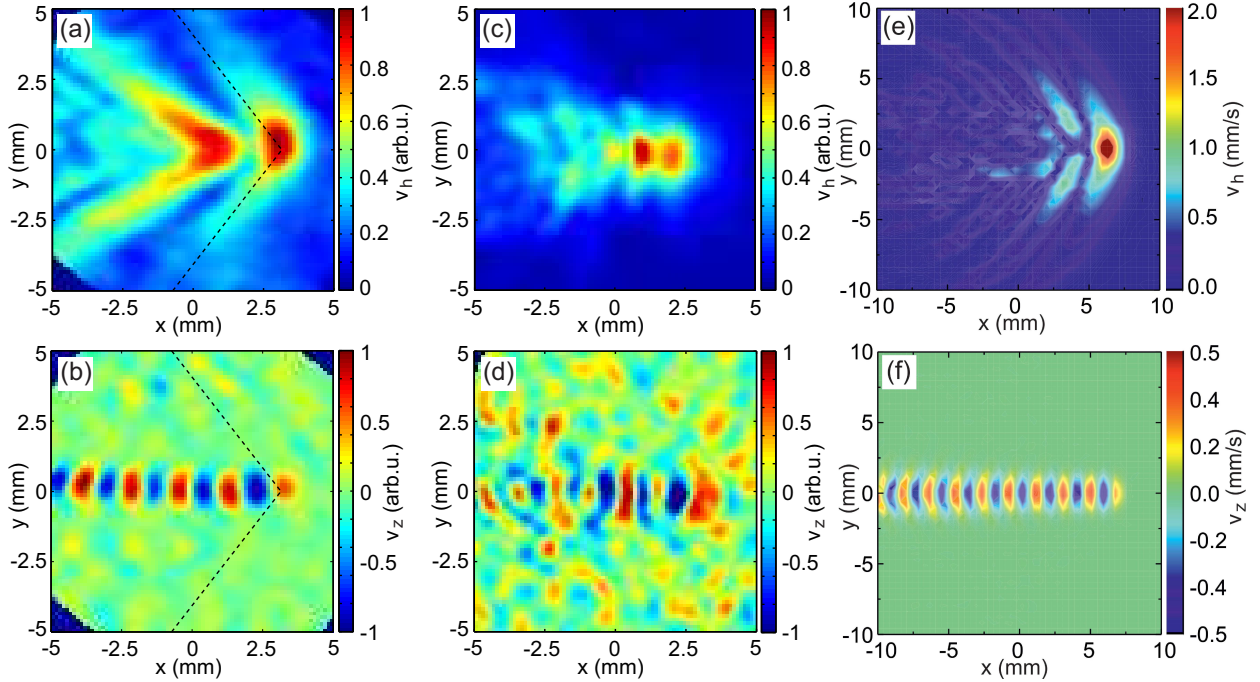}
\caption{Velocity maps visualising experiment I (a,b), experiment II
(c,d), and simulation (e,f), see Tab.\ref{tab1} for parameters. The
maps show the absolute value of the particle horizontal velocity $v_h$
(a,c,e) and the particle vertical velocity $v_z$ (b,d,f). The black
dashed line in (a,b) represents the theoretical longitudinal Mach
cone. The perturbation moved from left to right in all cases.
A compressional Mach cone with multiple structure is clearly visible
in cases, where the longitudinal Mach number was $>1$ (a,e). The
vertical wake structure (b,d,f) shows the vertical transverse waves
generated at the cone vertex by the perturbation.  The maps
were obtained by aligning consecutive frames at the cone vertex and
averaging them in order to reduce noise, 25 frames were averaged for
experiment I (a,b), 13 frames -- for experiment II (c,d), and 10
frames were used for simulation (e,f). 
\label{fig1_cones}
}
\end{figure*}

Both experimentally obtained (Fig.~\ref{fig1_cones}a-d) and simulated
(Fig.~\ref{fig1_cones}e,f) Mach cones have been visualised using the
method of Refs.~\cite{Samsonov1999b,Samsonov2000}. Several consecutive
frames have been aligned at the cone vertex and averaged, producing
high resolution velocity maps (Fig.~\ref{fig1_cones}). This reduced
the thermal noise and enhanced the motion correlated with the Mach
cone.  The horizontal speed is shown in Figs.~\ref{fig1_cones}a,c,e, it
visualises compressional Mach cones (Figs.~\ref{fig1_cones}a,e).  Note
that there is no compressional cone in Fig.~\ref{fig1_cones}c, since
the perturbing particle was subsonic. A double cone is visible in
Fig.~\ref{fig1_cones}a. The front compressional cone agreed well with
the theoretical prediction (dashed line). The second cone with a
smaller angle is a rarefactional cone \cite{Samsonov2000}.

Since the particles in a \textcolor{black}{quasi-2D} complex plasma are confined vertically in
a finite harmonic potential, they can move in the vertical direction,
 e.g. out of the lattice plane. A fast charged particle moving under the
main lattice layer exerts a 3D force and therefore perturbs the
lattice vertically as well as horizontally. While large scale vertical
lattice oscillations ~\cite{Piel2002} and waves
\cite{Samsonov2005,Couedel2009a} have been observed before, here a
long and narrow stripe of lattice is excited
vertically. Figures~\ref{fig1_cones}b,d,f show the vertical velocity
maps of lattices in the wake of an out-of-plane perturbing
particle. \textcolor{black}{It is worth noting (to avoid misunderstanding) 
that a vertical velocity component 
mapping on the $x,y$ plane is convenient way to reveal the vertical component of a Mach cone in a 
monolayer complex plasma. The maximal vertical particle displacements observed remained 
relatively small and, of course, not enough to reconstruct precisely the vertical 
structure of a Mach cone (as done in the experiments with 
3D complex plasma clouds \cite{Schwabe2011}).}
Both experiments and the simulation produce the same
vertical excitation pattern, with a wavelength close to $3a$ and a
width of the excited stripe of $3-5a$ \textcolor{black}{where $a$ is 
the average interparticle distance}. This width remained nearly
constant at different distances from the cone vertex indicating that
the VTW speed is so low, that the vertical wave is damped by the
neutral drag before it spreads noticeably. The theory-predicted VTW
speed was a small percentage of the compressional dust-lattice wave speed.
Interestingly the vertical oscillations persisted further away from
the cone vertex, than the compressional or shear waves, indicating
that they experienced less damping (Fig.~\ref{fig1_cones}b,d,f).

A linear analytical theory has been used to describe the vertical
lattice oscillations.  Since the strength of vertical confinement of a
stable monolayer is normally about 50 times higher than that
of horizontal confinement, the amplitude of the vertical oscillations
is much smaller than that of the horizontal oscillations, given that
the energy is evenly distributed over the oscillation modes
\cite{melzer01:_normal}. The amplitude of the vertical motion is also
much smaller than the lattice constant. The vertical displacement 
$z_{\mathbf{\Delta}}$  of the particles in a lattice obeys:
\begin{eqnarray}\label{eq1}
    \partial_{tt}z_{\mathbf{\Delta}}+\gamma \partial_{t}z_{\mathbf{\Delta}}+
    [\Omega_v^2 z_{\mathbf{\Delta}}-2\Omega_c^2\sum_{\Delta'}\widehat{\alpha}_{\mathbf{\Delta}
     -\mathbf{\Delta}'}z_{\mathbf{\Delta}'}]=\nonumber \\
    =f(\mathbf{\Delta}-\mathbf{V}t)
\end{eqnarray}
where $\mathbf{\Delta}=m\mathbf{a}_1+n\mathbf{a}_2$ ($m$ and $n$
are integers, $\mathbf{a}_{1,2}$ are the primitive translation vectors
\cite{Kittel1961}), the prime denotes the neighboring particles, 
$\gamma$ is the damping rate, $\Omega_v=2\pi f_v$
is the vertical confinement parameter,
$\displaystyle\Omega_c^2=\frac{Z^2e^2}{m_d\lambda_D^3}$ is the 
dust lattice frequency, $\widehat{\alpha}$ is the
dispersion operator (with the Fourier spectrum $\alpha_{\mathbf{k}}$
\cite{Couedel2011,Zhdanov2009}), and $f$ is the excitation force due
to a repulsive projectile moving with a velocity $\mathbf{V}$.
The projectile is a point charge moving at a height $z_p$ below
the lattice and interacting with it via a force $\displaystyle f_{\mathbf{k}}\propto
\exp(-\frac{|z_p|}{\Delta}\sqrt{\kappa^2+({\mathbf{k}}\mathbf{\Delta})^2})$.

\begin{figure}[htbp]
\centering
\includegraphics[width=\columnwidth,angle=0]{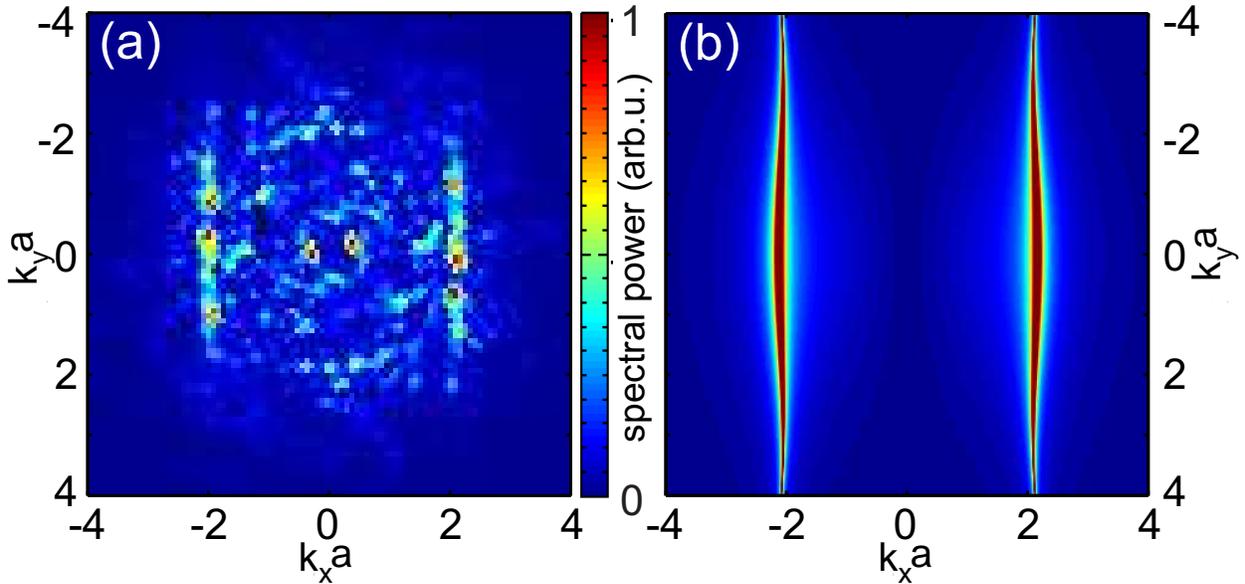}
\caption{2D Fourier transform of the vertical velocity map obtained in
experiment I, averaged over 10 frames (a), and its theoretical prediction
(b).}\label{fig_FFT}
\end{figure}
The solution of (\ref{eq1}) is a convolution of the exciter Fourier
spectrum $f_{\mathbf{k}}$, and the resonance propagator
$G_{\mathbf{k}}$ where the latter depends on
$\gamma$:
\begin{eqnarray}\label{eq2}
    z_{\mathbf{k}}=e^{i\mathbf{k}\mathbf{V}t}G_{\mathbf{k}}f_{\mathbf{k}},\\
    G_{\mathbf{k}}=[\Omega_{\mathbf{k}}^2-
    (\mathbf{k}\mathbf{V}+i\gamma)\mathbf{k}\mathbf{V}]^{-1},\\
    \Omega_{\mathbf{k}}^2=\Omega_v^2
    -2\Omega_c^2\alpha_{\mathbf{k}}.
\end{eqnarray}
The Fourier-transformed spectral intensity
of the vertical oscillations is, $I_{\mathbf{k}}\propto
\mathbf{v}_{\mathbf{k}}\mathbf{\overline{v}}_{\mathbf{k}}$,
$\mathbf{v}_{\mathbf{k}}=\partial_t z_{\mathbf{k}}$. In our case of low damping, it shows a
well-pronounced signature -- a narrow stripe transverse to the
direction of the projectile propagation in both experiment and theory
(Fig.\ref{fig_FFT}; in Fig.\ref{fig_FFT}(a), scattered bright dots, not well 
aligned vertically, are most probably the traces of previous
 excitations with different directions of propagation). The position of the stripe corresponds to the
resonance condition $\mathbf{V}_{ph}=\mathbf{V}$, where
$\mathbf{V}_{ph}$ is the vertical wave phase velocity. Therefore the
vertical structure of the Mach cone can be used to determine the
resonant wave number. It was found to be $ka=2.02\pm0.08$ in the
experiment, which agrees with the theoretical value of $ka=2.12$.  The
spatial decrement of the resonant vertical oscillations (\ref{eq2}) is $L^{-1}\simeq
\gamma/V$. 
Since the projectile velocity $V$ is relatively large, the spatial
decrement is low, and the wave
pattern persists for a long distance $\sim L$ explaining the seemingly lower
vertical damping (in experiment I, $L\sim$60 mm).

In conclusion, we have reported the first direct observation of the vertical \textcolor{black}{velocity}
component of a Mach cone in a quasi-2D complex plasma crystal excited by
a particle moving underneath the main layer. The vertical structure or z-Mach
cone was present in both cases of a super- or sub-sonic projectile. It is explained
by excitation of the vertical transverse wave in the wake of the projectile.
z-Mach cones can be used to determine the resonance wavelength of the vertical 
transverse wave.

\begin{acknowledgments}
The authors would like to thank D. Escande for useful comments and suggestions.
We appreciate funding from the European Research Council (Grant
agreement 267499), and from the Engineering and Physical Sciences
Reseach Council of the United Kingdom (Grant EP/G007918).
\end{acknowledgments}

\bibliographystyle{apsrev4-1}
%

\end{document}